# Two-dimensional MnB$_X$ monolayer as promising anode with record-high capacity and fast diffusion for Na-ion battery†

Yue Kuai[a], Changcheng Chen[a,*], Pengfei Lu[c]

*a School of Science, Xi'an University of Architecture and Technology, Xi'an, 710055, China.*

*b CAS Key Laboratory for Biomedical Effects of Nanomaterials and Nanosafety, Institute of High Energy Physics, Chinese Academy of Sciences, Beijing 100049, P. R. China*

*c State Key Laboratory of Information Photonics and Optical Communications, Beijing University of Posts and Telecommunications, Beijing, 100876, China.*

Email: chenchangcheng@xauat.edu.cn(C. Chen), wuly@bupt.edu.cn(L. Wu)

## ABSTRACT

Based on the group structure search method of first principles, MnB, MnB$_2$, and MnB$_6$ monolayer two-dimensional systems were designed and their structure, stability, electronic properties, as well as performance as anodes for sodium-ion batteries were examined. The findings indicate that the MnB, MnB$_2$, and MnB$_6$ monolayer two-dimensional systems possess high thermal stability, mechanical stability, dynamic stability, and distinctive metallic properties. When utilized in sodium-ion batteries, these MnB, MnB$_2$, and MnB$_6$ monolayers demonstrate high storage capacity, low diffusion barriers, and moderate open-circuit voltage. Due to its unique structure, the MnB monolayer presents a negative Poisson's ratio and a very low diffusion barrier. The theoretical capacitance of the MnB$_2$ monolayer, when acting as an anode for sodium-ion batteries, exceeds that of the MnB and MnB$_6$ monolayers. The research results reveal that boron-rich two-dimensional electrode materials pave a new way for energy applications. Particularly, MnB$_2$ and MnB$_6$ monolayer two-dimensional systems, as anode materials, incorporate the advantages of both planar and corrugated monolayer structures.

## 1. INTRODUCTION

Two-dimensional (2D) boron compounds, such as borophene, have been successfully fabricated on gold (111) substrates using molecular beam epitaxy [1-5]. This experimental success has opened up a new avenue for researching boron-based anode materials for sodium-ion batteries. Borophene exhibits high electrical conductivity, thermal conductivity, and excellent mechanical flexibility. In comparison to bulk boron, borophene can adopt various crystal structures due to the influence of different substrates, fabrication conditions, and interfacial electron transfer [6-10]. Moreover, the high specific surface area of borophene significantly increases the active centers for sodium storage [11-13].

First-principles calculations have demonstrated that different configurations of borophene can provide high theoretical capacities (504 and 1380 mAh/g) for sodium-ion batteries [14-18]. Two-dimensional monolayers composed of boron atoms and other elements also exhibit high sodium-ion storage capacities. For example, o-B2N2 (2159.83 mAh/g), B7P2 (3117 mAh/g), and BGe (1927 mAh/g) monolayers possess high theoretical capacities. However, similar to planar borophene monolayers, boron-based monolayers face the challenge of high diffusion barriers, hindering their application in metal-ion batteries [19-23]. Manganese-doped systems generally have lower diffusion barriers, which can be attributed to the lower spatial hindrance of manganese atoms [24-27]. Therefore, an effective strategy is to design monolayers formed by manganese and different proportions of boron atoms.

In this work, we systematically explore the potential of MnBx (x=1, 2, 6) monolayers as anode materials for sodium-ion batteries. MnBx (x=1, 2, 6) monolayers exhibit high stability, providing theoretical guidance for experimental fabrication. These monolayers maintain their metallic nature under various sodium-ion adsorption concentrations [28-32]. MnBx (x=1, 2, 6) monolayers also possess low sodium diffusion barriers and low average open-circuit voltages. The MnB2 monolayer achieves a capacity of 1050.37 mAh/g when applied to sodium-ion batteries, while the diffusion barrier for MnB monolayers is as low as 0.0017 eV. Sodium ions exhibit the lowest diffusion barrier in MnB monolayers. All these characteristics make MnBx (x=1, 2, 6) monolayers promising 2D high-capacity, low-diffusion-barrier materials for application as anode materials in sodium-ion batteries [33-38].

## 2. COMPUTATIONAL DETAILS

We performed the first-principles calculations using Vienna ab initio simulation package (VASP) [39,40]. The projector augmented wave (PAW) method [41] was also adopted in the calculation process. Perdew-Burke-Ernzerh (PBE) was used to describe the exchange-correlation functionals in the generalized gradient approximation (GGA) [42]. The cutoff energy of plane wave was set to 400 eV. The total energy and force converge to $10^{-5}$ eV and 0.01 eV, respectively [43]. In order to avoid the potential interaction between SiP₃ monolayers, a vacuum layer with a height of 20 Å was established in the vertical direction. During the optimizing geometry and static energy calculations, Monkhorst-Pack k-point was chosen as 3 × 3 × 1. The charge transfer between Li, Na, and K ions and MnB$_x$ substrates was analyzed by Bader charge method [44]. Phonon dispersion spectra were calculated using a 2×2 supercell and k-mesh of 16×16×1, using Phonopy code. The diffusion pathways of Li, Na, and K ions were calculated, and the energy of diffusion barrier was obtained by CI-NEB method [45]. The formation energy ($E_f$) of a compound can be calculated from the following equation:

$$E_f = \frac{E(Mn) + xE(B) - E(MnB_x)}{(x+1)} \tag{1}$$

Where $E(MnB_x)$ are total energy of MnB$_x$ monolayer, $E(Mn)$ and $E(B)$ are total energy of Mn atom and B atom in Bulk Structure. The positive formation energy represent the stable structure.

The adsorption energy ($E_{ad}$) of the alkali atom on MnB$_x$ monolayer can be obtained by [35]:

$$E_{ad} = \frac{E_{MnB_x} + nE_{Na} - E_{Na@MnB_x}}{n} \tag{2}$$

Here, the charging/discharging processes of MnB$_X$ monolayer were simplified as the half-cell reactions below

$$nNa^+ + xe^- + MnB_x \leftrightarrow Na_n@MnB_x \tag{3}$$

According to the above reaction model, we can get the following formula to calculate open-circuit voltage ($V_{OCV}$):

$$V_{OCV} = \frac{E_{MnB_x} + nE_{Na} - E_{Na_nMnB_x}}{ne} \tag{4}$$

where $E_{Na_nMnB_x}$ is the total energy of Na ion adsorbed on MnB$_X$ monolayer, 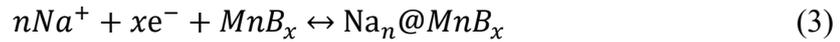 $E_{MnB_x}$ represents the energy of the pristine monolayer, $E_{Na}$ refers to the energy of a Na atom in the bulk system, and $x$ indicates the number of Na atoms [46]. The negative values of adsorption energy indicate that the Na atom is adsorbed on the pristine monolayer.

The maximum storage capacity $C$ is given by the following equation :

$$C = \frac{n_{max} \times F}{M_{MnB_x}} \tag{8}$$

where $M_{MnB_x}$ represents the atomic mass of 2×2 MnB$_x$ supercell, $n_{max}$ is the number of Na adsorbed at the maximum adsorption concentration, and $F$ is the Faraday constant (26.81 A h mol⁻¹).

The lattice change $\Delta x$ is expressed by the following formula:
$$\Delta x = \frac{a_n - a_0}{a_0} \tag{9}$$
where $a_n$ is the MnB$_x$ monolayer lattice constants of the maximum Na adsorption concentration. $a_0$ is the pristine MnB$_x$ monolayer lattice constants [47].

The temperature-dependent diffusivity of MnB$_x$ monolayer was evaluated using the Arrhenius equation defined as following:
$$D = l^2 v_0 exp\left(\frac{-E_a}{K_B T}\right) \tag{10}$$
where $l$ is the migration distance from equivalent Na adsorption sites on the MnB$_x$ monolayer, $K_B$ is the Boltzmann constant, $E_a$ is the diffusion barrier, $T$ is the absolute temperature and $v_0$ is the vibration frequency [48-50].

## 3. RESULTS AND DISCUSSION

### 3.1 Geometry, electronic properties, and stability of MnB$_x$(x=1, 2, 3) monolayer

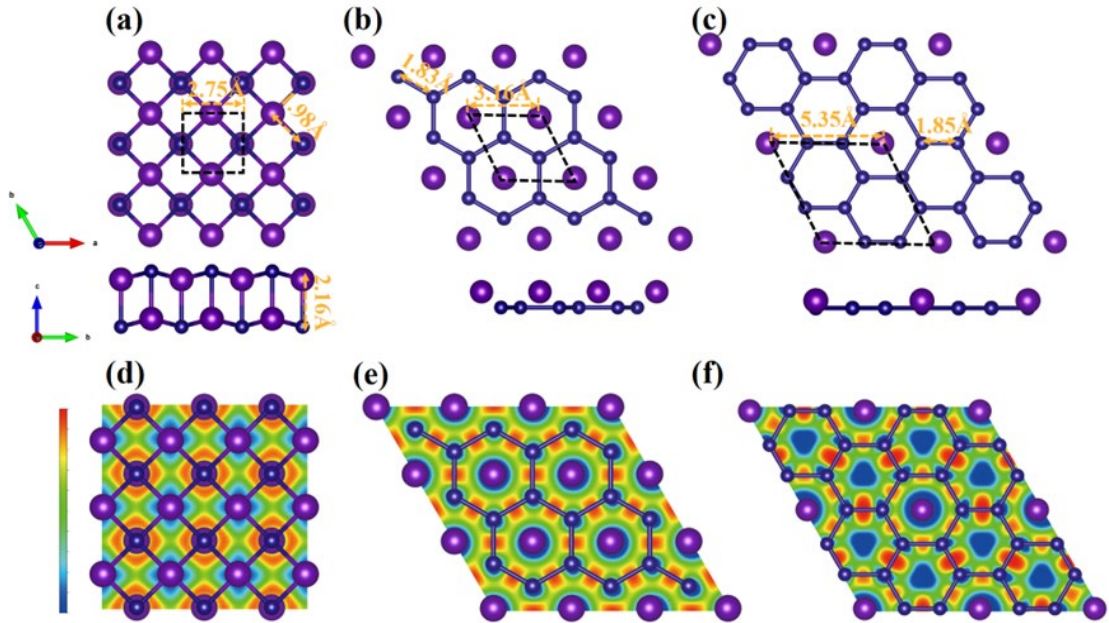

**Fig. 1.** Optimized structural of (a) MnB, (b) MnB$_2$ and (c) MnB$_6$ monolayer, deepblue spheres represent B atoms, purple spheres are Mn atoms. Electron localization function of the (001) section of (d) pristine MnB, (e) MnB$_2$ and (f) MnB$_6$ monolayer. Increasing electron localization from 0 to 1 is plotted with colours from dark blue to red.

Three MnB$_x$(x=1,2,6) monolayer with different boron ratios were predicted by global particle swarm optimization method using CALYPSO code: MnB, MnB$_2$ and MnB$_6$. As shown in **Fig. 1**, these MnB, MnB$_2$ and MnB$_6$ monolayer are all formed by coordination between manganese atoms and surrounding boron atoms. In the predicted MnB structure (**Fig. 1a**), each manganese atom is coordinated with four adjacent boron atoms, in which the MnB monolayer is a buckling structure [51-53]. The lattice constant of the primitive cell, the B-Mn bond length in the plane and the bond length along the c-axis are 1.98 and 2.16 Å, respectively.

In the $MnB_2$ monolayer, the lattice constant of $MnB_2$ (**Fig. 1 a**) and the bond length of B-B are 1.83 Å. The unit rings of $MnB_6$ monolayer are adjacent to each other and share two boron atoms with six adjacent units. Fig. 1 b shows the optimized structure of the $MnB_2$ monolayer [54]. The lattice constants a and b of the print cell are 3.16 Å, and the bond lengths of B-B are 1.99 and 1.74 Å, respectively. As shown in Fig.1 c, the lattice constant of $MnB_6$ is a=b=5.35 Å, and the bond length of B-B is 1.85 Å. The isolated $MnB_6$ units are connected by the B-Mn-B bond and arranged periodically. Due to of the high atomic density of manganese and boron, the single-layer of MnB, $MnB_2$ and $MnB_6$ have quasi-plane two-dimensional geometric structure. The buckling geometry helps to avoid the repulsive interaction between high-density manganese and boron atoms, thus reducing the total energy of the monolayer.

In order to further understand the structural properties, we have drawn the electronic local function (ELF) of MnB, $MnB_2$ and $MnB_6$ monolayer in Fig. 1d-f. The ELF values of 0 indicate complete delocalization of electrons and 0.5 means the uniform electron gas and 1 indicate complete localization [55,56]. In MnB monolayer (Fig. 1d), dense charge localization is shown around the B atom. As shown in Fig. 1e, the ELF values of between B atoms is ~0.87 in $MnB_2$ monolayer, which means the formation of covalent bond between B atoms. In $MnB_6$ monolayer, B-B bond is also covalent bond. As shown in Fig. 1d and Fig. 1f, the $MnB_6$ monolayer is different from $MnB_2$ monolayer in ELF between B atoms. The ELF values of all B-B bonds are not equal, only the ELF values in the B ring embedded by Mn atom are equal, which is similar to that of $MnB_2$ monolayer. Around the Mn atoms in MnB, $MnB_2$ and $MnB_6$ monolayer, the corresponding ELF value is near to 0, representing the depletion of the electron. As shown in Table S1. The results of Bader analysis revealed that in the MnB, $MnB_2$ and $MnB_6$ monolayers 0.4, 0.38 and 0.34 electrons were transferred from the d orbital of each Mn atom to the B atom, respectively.

The dynamic stability of the MnB$_x$ (x=1,2,6) monolayer is evaluated by calculating the phonon spectrum along the high symmetry line in the Brillouin zone. As shown in Fig. 2 (a-c), there is no obvious imaginary frequency in the phonon spectrum of MnB$_x$ (x=1,2,6) monolayer, which indicates that the MnB$_x$ (x=1,2,6) monolayer has dynamic stability. The highest frequencies of MnB, MnB$_2$ and MnB$_6$ monolayer are 721.53 cm$^{-1}$, 805.92 cm$^{-1}$ and 1171.43 cm$^{-1}$, respectively. The high phonon spectrum frequency shows that there is a strong interaction between Mn-B and B-B in the MnB$_x$ monolayer, and the frequency of MnB$_6$ up to 1171.43 cm$^{-1}$ is mainly attributed to the tensile vibration contribution of the B-B bond [58].

AIMD simulations were carried out to evaluate the thermodynamic stability of MnBx monolayer at 500K. The relationship between time and total energy function of ab initio molecular dynamics simulation is shown in Fig .2(d-f). After simulating 1ps, the energy tends to converge and oscillates in a very small range [59,60]. This proves that the single-layer of MnB$_x$ has thermodynamic stability at 500K.

Young's modulus reflects the flexibility and rigidity of the material, and Poisson's ratio reflects the mechanical response of the material to external load, which is the two main mechanical parameters of the material. The electrode material is subject to repeated expansion during charging and discharging [61]. The mechanical properties of the MnB$_x$(x=1,2,6) monolayer are therefore closely related to the cycling stability of the battery. According to the Born elastic stability standard, the elastic constants of the have to satisfy $C_{11}$, $C_{22}$, $C_{44}$>0 and $C_{11}C_{22}$>$C_{12}^2$. The elastic constants of the MnB$_x$(x=1,2,6) monolayers satisfy the Born elastic stability standard, indicating that the MnB$_x$(x=1,2,6) monolayers are mechanically stable. As shown in Figure 5.2(g-h), the MnB, MnB2 and MnB6 monolayers exhibit maximum Young's modulus of 120.52, 100.48 and 128.33 N/m respectively.

The thermodynamic stability of the MnB$_x$(x=1, 2, 6) monolayer was assessed by calculating and analysing the cohesion and formation energies of the predicted MnB$_x$(x=1, 2, 6) monolayer in order to assess the feasibility of its experimental preparation, with negative cohesion energies implying the energy released when isolated free atoms form the compound [62,63]. Cohesion energies of the MnB, MnB$_2$ and MnB$_6$ monolayer are -5.53 eV, -5.68 eV and -5.71 eV respectively. the relatively large cohesion energies of the $MnB_x$(x=1, 2, 6) monolayer compounds therefore indicate that the monolayers are promising for experimental preparation. Moreover, we assess the energy change of a compound when it is formed from its constituent elements under its monomer by calculating the formation energy. This is attributed to two factors: i) In general, isolated atoms are not a realistic alternate phase and compounds are rarely synthesised from isolated atoms. Therefore, the cohesion energy is not a suitable measure. ii) The phonon calculation and AIMD simulations for $MnB_x$(x=1, 2, 6) monolayer show that there is at least some kinetic barrier slowing decomposition, but for long-term operation in a battery it's important to know whether there is a thermodynamic driving force for MnBx(x=1, 2, 6) monolayer to turn into one or more other phases. Therefore, the formation energy was calculated to assess the thermal stability of the MnBx(x=1, 2, 6) monolayer to determine whether the $MnB_x$(x=1, 2, 6) monolayer is susceptible to decomposition. As shown in Fig. 2(i), the formation energies of the MnB, $MnB_2$ and $MnB_6$ monolayers are -0.25, -0.33 and -0.40 eV respectively [64]. The negative formation energies indicate that the $MnB_x$(x=1, 2, 6) monolayers are thermally stable. In addition, the thermal stability gradually increases with increasing boron ratio.

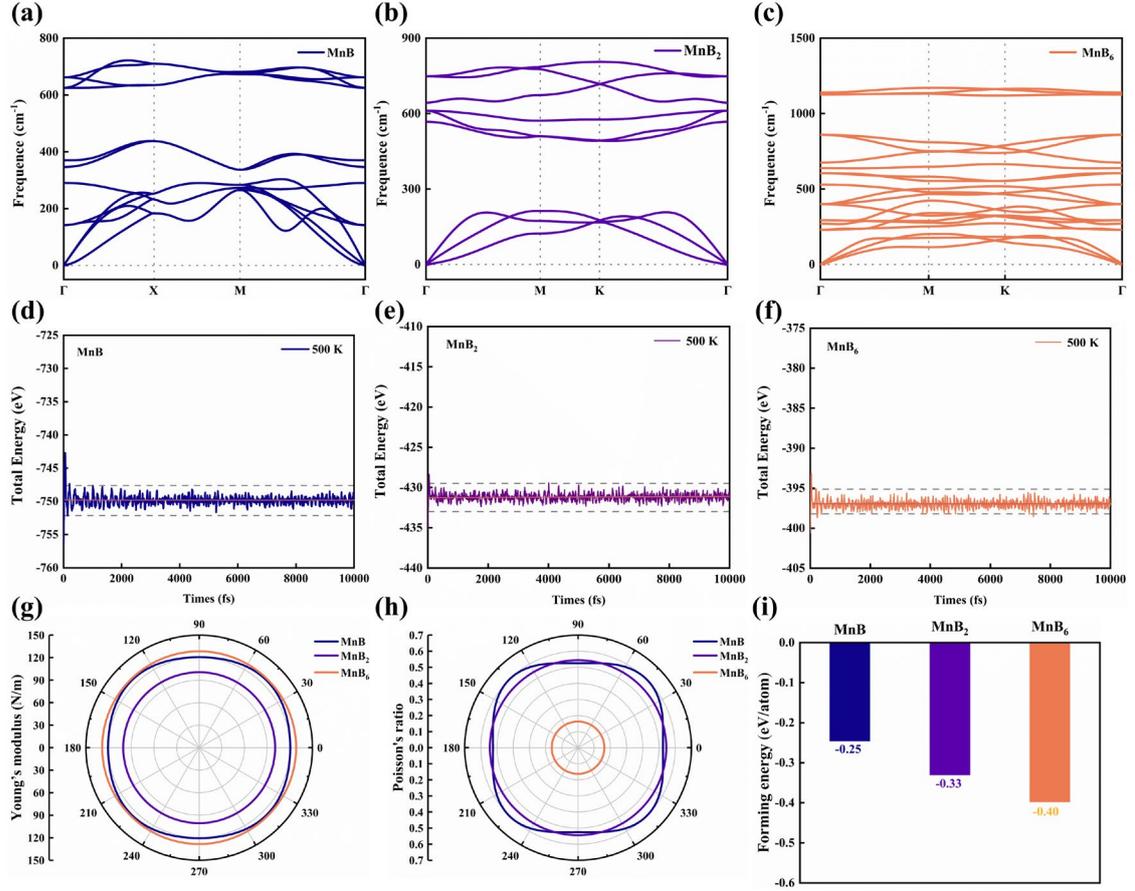

**Fig. 2.** (a) phonon spectra of MnB, (b) MnB2 and (c) MnB6 monolayer. AIMD simulations of energy oscillation diagrams for supercell (d) MnB, (e) MnB2 and (f) MnB6. (g) Young's modulus and (h) Poisson's ratio of MnB, MnB2 and MnB6 monolayer. (I) the formation energy of MnB, MnB2 and MnB6 monolayer.

As shown in Fig. 3, the electronic properties of the single-layer of MnBx include energy band, total density of states and projected density of states. Usually, the band gap of the material is underestimated by using PBE functional, and the band gap is overestimated by HSE06 functional [65]. Because manganese is a magnetic atom, spin orbit coupling needs to be taken into account, so HSE06+SOC is used to calculate the energy band of $MnB_x$ monolayer. Fig. 3 the band and the projected density of states diagram show that the MnB, $MnB_2$ and $MnB_6$ monolayer show unique metal properties, and their valence bands pass through the Fermi level. The valence band and the high density of states peak pass through the Fermi level, which confirms the electronic conductivity. The projected density of states in Fig. 3 (d-f) shows a high density of states at the Fermi level, which is mainly due to the contribution of the d orbitals of the Mn atom.

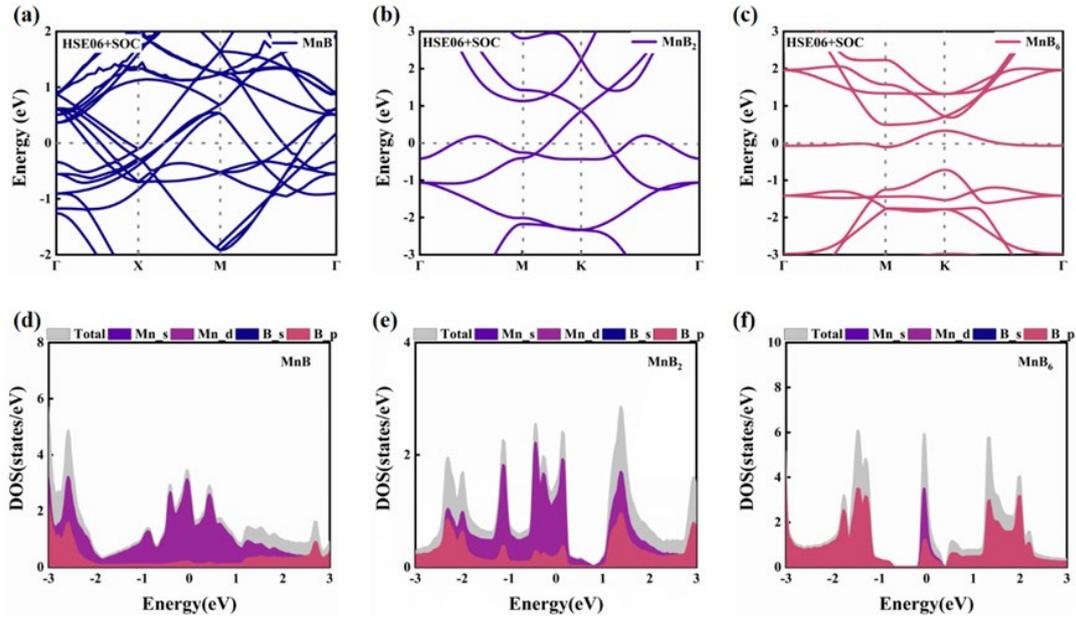

**Fig. 3.** The band structures and PDOS of (a) and (d) MnB monolayer, (b) and (e) MnB2 monolayer, (c) and (f) MnB6 monolayer. The fermi-level is set at zero.

**3.2 Single Na ion adsorption and electronic properties**

Understanding the storage properties of Na ions in MnB$_x$ (x=1,2,6) monolayers requires a crucial investigation into the adsorption of single Na ions (Li, Na, K) on MnBx (x=1, 2, 3) monolayers. The adsorption properties of Na ions were analyzed by calculating and analyzing the adsorption energy of the adsorption site, differential charge density, and Bader charge [66,67]. The variety of Na ion adsorption sites is determined by the unique geometric structure of MnB, MnB2, and MnB6 monolayers. In the MnB monolayer two-dimensional, based on the geometric structure of MnB monolayer with ab-plane mirror symmetry, the potential adsorption sites of sodium ions on MnB monolayer, including manganese and boron atoms, vacancy centers on the layer, and Mn-B bonds, labeled as M1, B1, H1, and L1, respectively, were investigated. By optimizing the structures of different adsorption sites, it was determined that M1, B1, and H1 are the stable adsorption sites for sodium ions. The Na ions adsorbed at L$_1$ site moved to M$_1$ site after relaxation [68]. In the MnB2 monolayer, the arrangement of Mn and B atoms is non-coplanar, as depicted in **Fig. 1b**. Potential Na ion adsorption sites include M2a/M2b, B2a/B2b, T2a/T2b, and L2a/L2b, which correspond to positions above and below the manganese atoms, above and below the boron atoms, above and below the B-B bonds, and in the middle above and below the Mn-B atoms, respectively. The computational analysis indicates that M2a/M2b, B2a/B2b, and T2a/T2b are stable adsorption sites. The geometric structure of MnB6 monolayer is similar to that of MnB2 monolayer. Therefore, the possible sodium ion adsorption sites are similar to those of MnB2, which are set as T3a/T3b, M3a/M3b, H3a/H3b, B3a/B3b, and L3a/L3b, above and below the manganese and boron atoms, at the center of the boron hexagon, above and below the B-B bond, and above and below the Mn-B bond, as shown in Figure 1c. Among them, the adsorption sites of manganese atoms, boron atoms, and boron hexagons are stable. The sodium ions adsorbed above and below the B-B bond will move to the center of the hexagon after structural optimization [69]. The sodium ions adsorbed in the middle of the B-Mn bond will move to the boron atom after structural optimization. Figure 1f shows the stable adsorption energy of MnBx (x=1,2,6) monolayers. The H3a site of MnB6 monolayer is the most stable sodium ion adsorption site.

Figure 4d-f show the adsorption energy profiles of a single Na ion at the stable adsorption sites in MnB, MnB2, and MnB6, respectively. We found that the most stable adsorption sites in these three monolayers are not located on the same atoms. The most stable adsorption sites are located on Mn atoms, B atoms, and B hexagonal rings, respectively, which are related to the surface structure and electron density distribution of the three materials. Na atoms seem to be more easily embedded into larger gaps. In MnB, Mn atoms are close to the inner edges of the unit cell, forming a concave surface. In MnB2, three Na ions are embedded in the gaps between surface Mn atoms and located above B atoms. MnB6 provides large hexagonal ring gaps. The common feature is that the stable adsorption sites are a concave, which provides larger space and stronger adsorption energy, making it easier for Na ions to adsorb. The hexagonal ring centers also have higher electron density and stronger chemical reactivity, which can form stronger chemical bonds with alkali metal ions.

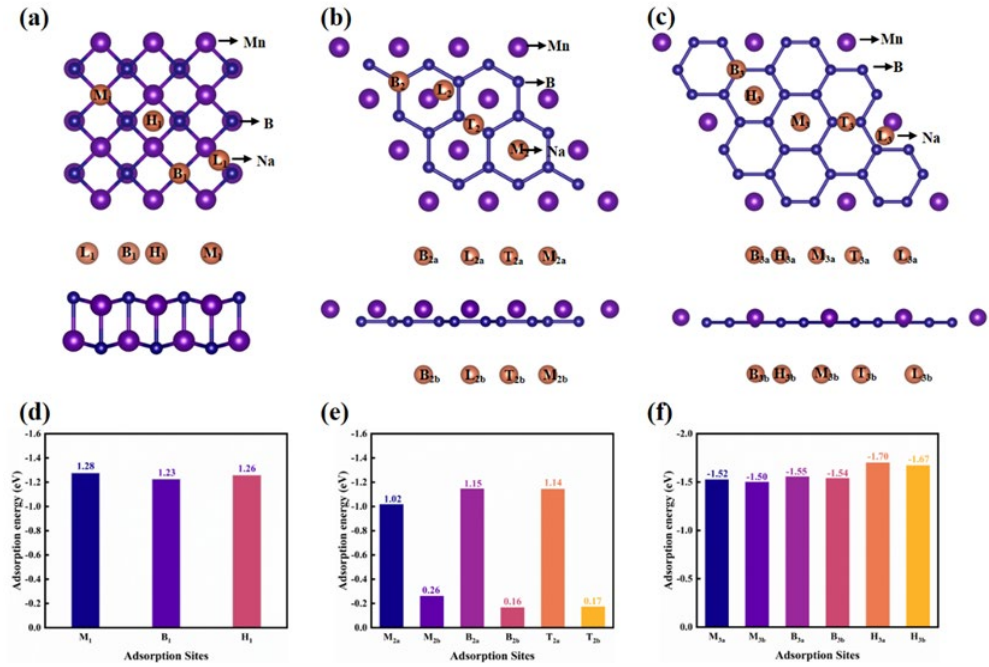

**Fig. 4.** The band structures and PDOS of single (a) and (d) Li ion, (b) and (e) Na ion, (c) and (f) K ion adsorbed on SiP₃ monolayer. The fermi-level is set at zero.

To further investigate the metal atom adsorption mechanism, the stable adsorption sites of MnBx (x = 1, 2, 6) monolayer were probed by differential and Bard charges. The differential charges of sodium ions at the most stable adsorption sites (M1, B2a, H3a) of MnB, MnB2 and MnB6 monolayer are shown in Figure 5.5(a-c). The blue color represents the loss of electrons and the yellow color represents the gain of electrons. Electron transfer from the sodium ion to the MnB, MnB2 and MnB6 monolayer can be clearly observed. Further analysis of the charge relative transfer by Bard's charge shows that the sodium ion transfers 0.775, 0.792 and 0.828 electrons to the MnB, MnB2 and MnB6 monolayer, respectively, in the stable adsorption site structure. This indicates that there is a significant electron transfer between sodium ions and the monolayer of MnB, MnB2 and MnB6, and the interaction between sodium ions and these metal atoms is mainly electron transfer to each other, rather than just by van der Waals forces or electrostatic interactions.

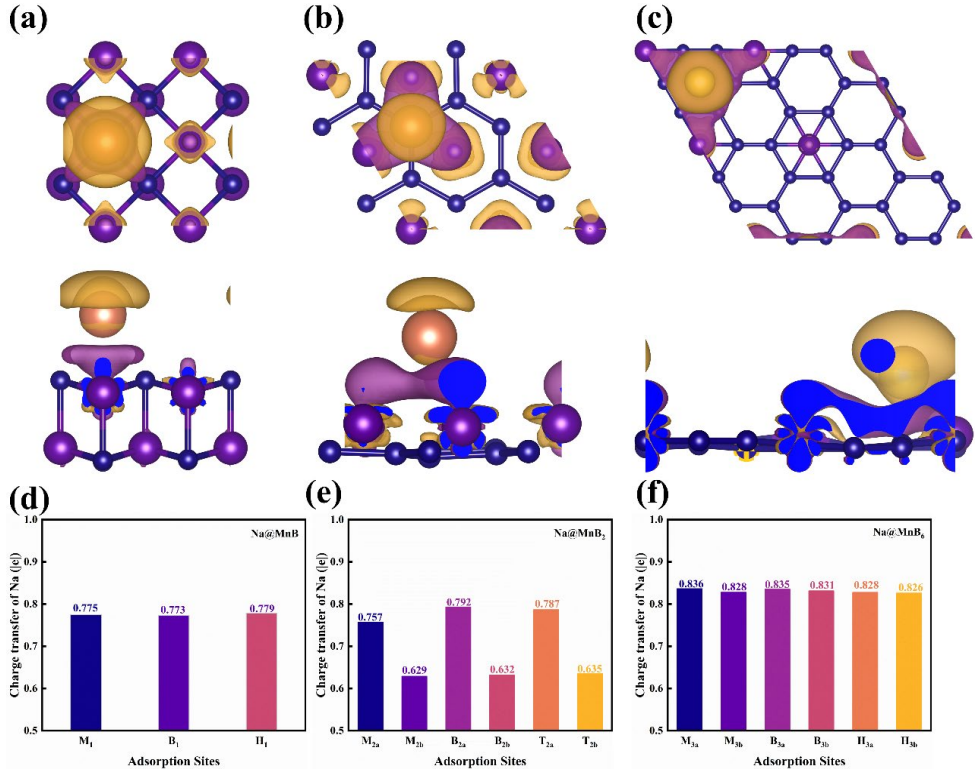

**Fig. 4.** Differential charge of sodium ions adsorbed at the M1 site of MnB monolayer (a), the B2a site of MnB2 monolayer (b), and the H3a site of MnB6 monolayer (c). Charge transfer of stable sites for sodium ion adsorption on (d) MnB, (e) MnB2, and (f) MnB6.

**3.3 average open-circuit voltage (OCV)**

The open-circuit voltage and the theoretical capacity are the key parameters to measure the performance of the battery, and an open-circuit voltage between 0-1 V can avoid the growth of metal clusters during the charging and discharging process. After determining that sodium ions can be adsorbed in the $MnB_x$(x=1, 2, 6) monolayer, the adsorption energy was first calculated and analyzed in relation to the adsorption concentration in order to further investigate the sodium storage performance of the $MnB_x$(x=1, 2, 6) monolayer, as well as the open-circuit voltage and theoretical capacity of the $MnB_x$(x=1, 2, 6) monolayer as the anode of sodium ion batteries [70]. In Figure 5.6(a-c), the adsorption energy gradually tends to 0 eV with increasing sodium ion concentration for MnB, MnB2 and MnB6 monolayer, implying that the adsorption concentration of sodium ions increases with increasing sodium ion adsorption leading to the repulsion between neighboring sodium ions and thus making the adsorption unstable. In order to determine the open circuit voltage distribution of $Na_nMnB_x$, in Figure 5. 6(d-f), the $Na_nMnB_x$ formation energy was calculated and constructed as energy profiles to evaluate the stability of the intermediate phase, with structures located on the solid line as stable phases and those above the convex packet as sub-stable phases. For the $Na_nMnB_x$ system, eight different sodium adsorption concentrations (n=0.125, 0.25, 0.375, 0.5, 0.75, 0.875 and 1) were considered, where the adsorption configurations corresponding to n of 0.125, 0.25, 0.375, 0.5 and 1 have good stability, and the adsorption configurations corresponding to concentrations n of 0.75 and 0.875 have were less stable. For the $Na_nMnB_2$ system, six different sodium ion concentrations (n=0.5, 1.0, 1.5, 2.0, 2.5 and 3.0) were considered, where n of 0.5, 1.0, 1.5, 2.5 and 3.0 corresponded to adsorption configurations with good stability. For the $MnB_6$ monolayer, the adsorption configurations corresponding to n of 0.5, 1.0, 1.5, 2.0, 2.5, 3.0, 3.5 and 4.0 have good stability among the eight adsorption concentrations considered (n = 0.5, 1.0, 1.5, 2.0 and 4.0). After determining the stability of $Na_nMnB$, $Na_nMnB_2$ and $Na_nMnB_6$ intermediate phases, the open-circuit voltage variation of $MnB_x$ (x = 1, 2, 6) monolayer as anode materials for sodium ion batteries is plotted in Figure 5.6(g-i). In the open-circuit voltage distribution, the interaction between sodium ions is weak at low ion concentrations, and the repulsion between sodium ions increases as the adsorption concentration increases, and strong repulsion occurs when the adsorption concentration is large. The open-circuit voltage values of these three systems decrease gradually with increasing concentration, and the average open-circuit voltages of the simulated $MnB_x$ (x = 1, 2, 6) monolayer are 0.79 V, 0.43 V and 0.76 V, respectively, all of which lie in the appropriate range and reduce the risk of dendrite formation.

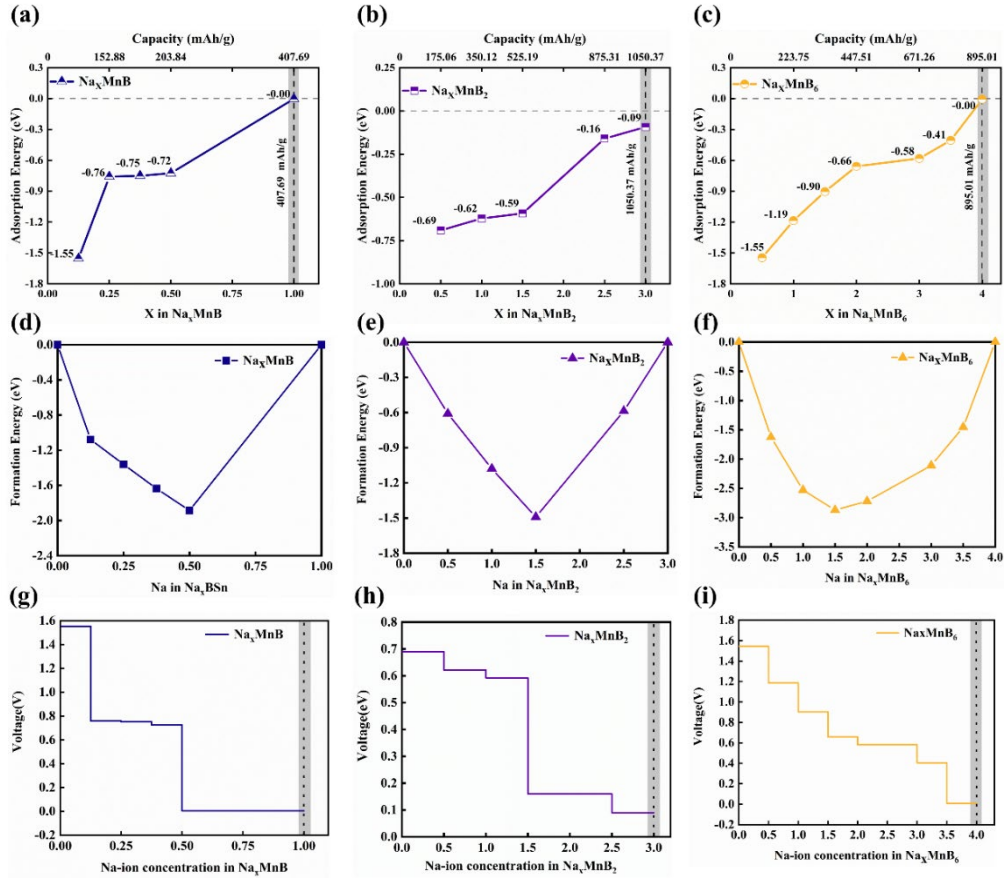

**Fig. 5.** The relationship between the adsorption energy and adsorption concentration of sodium ions in (a) MnB, (b) MnB$_2$, and (c) MnB$_6$ monolayer. The corresponding formation energy of sodium ion concentration adsorbed on (d) MnB, (e) MnB$_2$, and (f) MnB$_6$ monolayer. Voltage dependence of sodium ion concentration on monolayer adsorbed on (g) MnB, (h) MnB$_2$, and (i) MnB$_6$.

Studying the diffusion kinetics of ions at the atomic scale inside batteries has guiding significance for evaluating the rate performance during battery charging and discharging processes. This chapter investigates the diffusion kinetics of MnB, MnB$_2$, and MnB$_6$ monolayers through ab initio molecular dynamics simulations and climbing micro elastic band methods. For the climbing micro elastic band method, it is necessary to explore suitable paths [71]. The potential barrier calculated by the climbing micro elastic band method is based on the structure at 0K temperature, but in actual battery operation, lattice thermal vibration is inevitable. Therefore, first calculate the molecular dynamics simulation of a single sodium ion on a MnB$_x$ (x=1,2,6) monolayer at a temperature of 300K. The motion trajectory of a single sodium ion in a MnB$_x$ (x=1,2,6) monolayer is shown in Figure 5.7. The molecular dynamics trajectory of sodium ions in the MnB monolayer at 5ps shows that sodium ions can pass through the entire MnB monolayer region. In the 1-2.7ps simulation process, sodium ions first diffuse along the adjacent H site (Mn-B central vacancy), then diffuse from the H1 site to the B1 (above B atom) site, and diffuse almost linearly along the B-Mn-B-Mn-B atom to the entire region. Subsequently, during the 2.7-5ps simulation process, sodium ions diffused in an approximate arc manner over the MnB monolayer, first diffusing between B-B atoms, then continuing to diffuse above B-Mn B-Mn atoms, and finally diffusing between Mn Mn atoms. Based on the above dynamic simulation trajectory, three representative diffusion paths were considered and the diffusion potential barrier was calculated. Path-1 is a diffusion pathway composed of four adjacent H sites [72]. Path-2 refers to the diffusion of sodium ions from the H site through the B site to another H site, along the diffusion path of the H-B-H site. Path-3 refers to the diffusion of the M site through the B site to another M site, which follows the diffusion path of the M-B-M site.

Molecular dynamics simulations of sodium ions in the $MnB_2$ monolayer were performed for 10 ps, and the sodium ion motion trajectory can still pass through the whole region. It takes longer time compared to the MnB monolayer. In 1-5.0 ps, the sodium ions first move between adjacent boron atoms and later diffuse through the B-Mn, Mn-Mn, and Mn-B-B-Mn paths above the atoms. Finally, in 5-10 ps, diffusion continues along the adjacent B-B atoms as well as between Mn-Mn atoms. Three representative diffusion paths are selected: Path-1: four adjacent $B_{2a}$ sites constitute; Path-2 is the diffusion of sodium ion from B2a site to another $B_{2a}$ site via $Mn_{2a}$ site; Path-3 is the diffusion of sodium ion between two adjacent $M_{2a}$ sites.

In the 10 ps molecular dynamics simulation of $MnB_6$, the sodium ion first diffuses in the $MnB_6$ monolayer two-dimensional system and then stabilizes in the boron hexa ring center site. In 1-5.0 ps, the sodium ion first diffuses from one manganese atom to another, and then the sodium ion diffuses from the manganese atom to the center of the boron hexa ring (H site) and continues to diffuse to another manganese atom. Within 5-10 ps, the sodium ion diffuses from above the manganese atom to another boron hexa ring center and then stably adsorbs in the boron hexa ring center. The selected representative diffusion paths are: Path-1 (Path-1) is the sodium ion diffusion path in three adjacent $H_{3a}$ sites; Path-2 (Path-2) is the sodium ion diffusion from $H_{3a}$ site through $M_{3a}$ site to another $H_{3a}$ site, both $H_{3a}$-$M_{3a}$-$M_{3a}$ diffusion path; Path-3 (Path-3) is the sodium ion diffusion from H3a site through $B_{3a}$ site to another $H_{3a}$ site, both the $H_{3a}$-$B_{3a}$-$H_{3a}$ pathway.

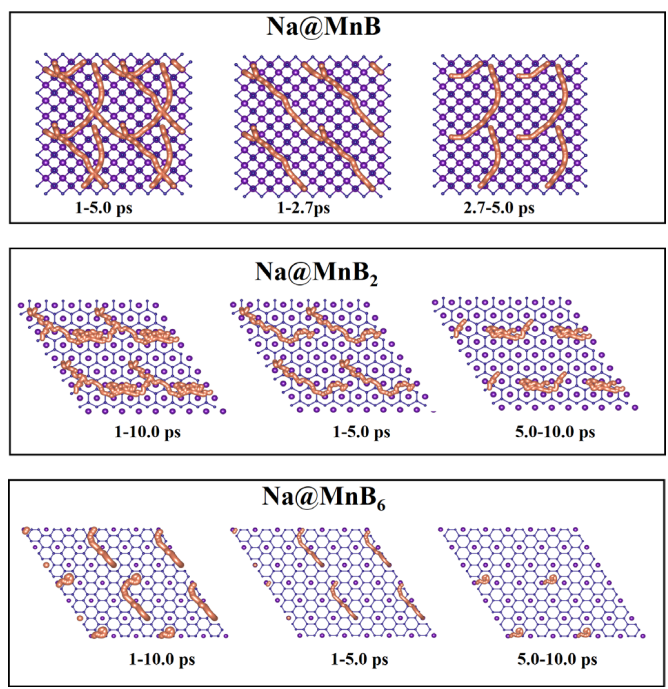

**Fig. 6.** AIMD simulation of diffusion trajectories of individual sodium ions in MnB, $MnB_2$ and $MnB_6$ monolayer at 300 K temperature conditions.

In anode materials, a lower diffusion barrier for metal ions indicates a higher charge and discharge rate for ion batteries, as well as a faster diffusion migration rate. The diffusion barriers of sodium ions along Path-1, Path-2, and Path-3 in the MnB, MnB2, and MnB6 monolayers are investigated. In Fig. 5.8(a-c), the diffusion barriers of sodium ions along Path-1, Path-2, and Path-3 in the MnB monolayer are found to be 0.0017 eV, 0.041 eV, and 0.046 eV, respectively, with a negligible barrier observed along Path-1. The diffusion barriers of sodium ions in the MnB2 monolayer are displayed in Fig. 5.8(d-f), with values of 0.058 eV, 0.01 eV, and 0.048 eV for Path-1, Path-2, and Path-3, respectively. Compared to the diffusion barrier of sodium ions along Path-1 in the MnB monolayer, the barrier along Path-2 in the MnB2 monolayer is 10 times higher. The diffusion barriers of sodium ions along Path-1, Path-2, and Path-3 in the MnB6 monolayer are presented in Fig. 5.8(g-i), with values of 0.129 eV, 0.126 eV, and 0.15 eV, respectively. Among the MnB, MnB2, and MnB6 monolayers, sodium ions exhibit an almost zero diffusion barrier in the MnB monolayer, while the MnB6 monolayer displays the highest diffusion barrier [73]. The relatively low diffusion barriers in the MnBx (x=1, 2, 6) monolayers can be primarily attributed to the lower steric hindrance of manganese atoms. As the number of boron atoms increases in the

MnBx (x=1, 2, 6) monolayers, the steric hindrance is enhanced, resulting in a higher diffusion barrier for the MnB6 system than for the MnB2 and MnB systems. Thus, the diffusion barrier hierarchy can be summarized as MnB6 > MnB2 > MnB.

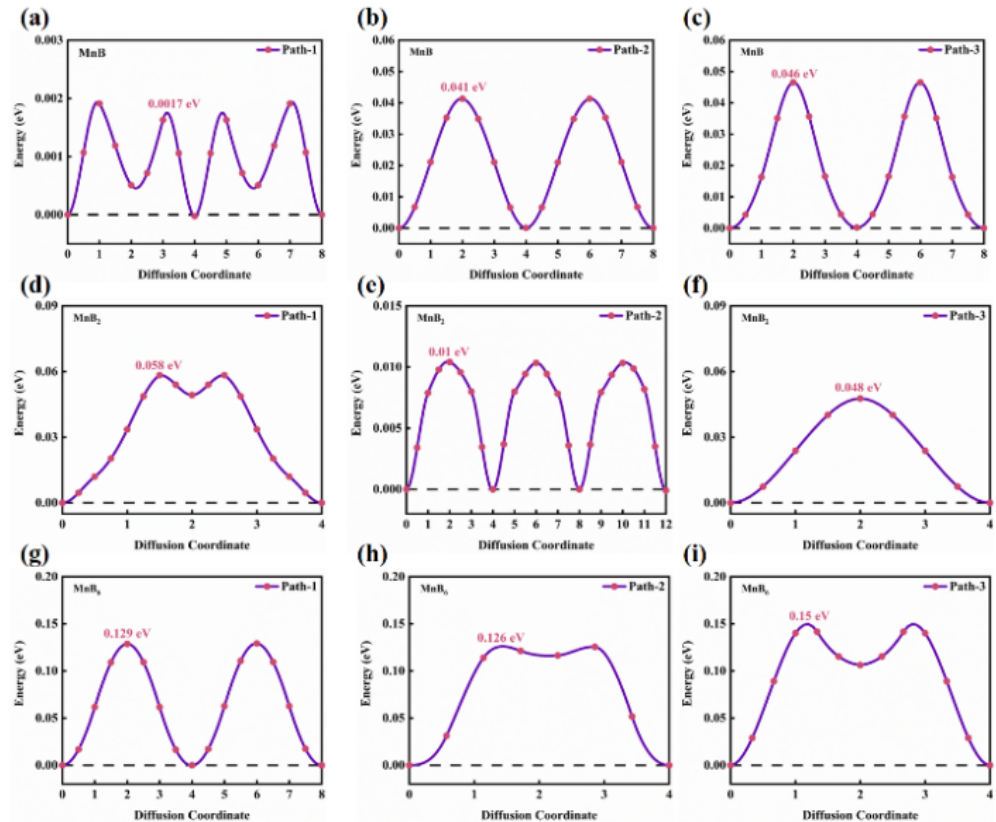

**Figure 5.7** shows the diffusion barriers for sodium ions along Path-1, Path-2, and Path-3 on monolayer two-dimensional systems of MnBx (x=1, 2, 6). Panels (a-c), (d-f), and (g-i) correspond to monolayer two-dimensional systems of MnB, MnB2, and MnB6, respectively.

## 4．CONCLUSION

In this study, we employed a particle swarm optimization algorithm combined with first-principles calculations to predict the properties of three kinetically, energetically, mechanically, and thermally stable monolayers of MnB, MnB2, and MnB6 by controlling the concentrations of manganese and boron atoms. Our findings reveal that the monolayers of MnB2 and MnB6 exhibit high capacities as anode materials for sodium-ion batteries (1050.37 mAh/g and 895.01 mAh/g, respectively). Additionally, the intrinsic metallic nature and excellent thermal stability of MnBx (x=1, 2, 6) monolayers provide good electronic conductivity and cycling performance during the intercalation/deintercalation process of sodium ions. The sodium-ion diffusion barriers for MnB, MnB2, and MnB6 monolayers are as low as 0.0017 eV, 0.01 eV, and 0.126 eV, ensuring rapid charging and discharging during battery operation. Furthermore, the low average open-circuit voltages for MnB, MnB2, and MnB6 monolayers as sodium-ion battery anodes are 0.79 V, 0.43 V, and 0.76 V, respectively, effectively preventing the formation of dendrites in the 0-1V range. Among these systems, the MnB2 monolayer demonstrates the most exceptional theoretical capacity, while the MnB monolayer exhibits the best rate performance.

■ **ASSOCIATED CONTENT**

**Supporting Information**

The Supporting Information is available free of charge at structural parameters of $XP_3$ monolayers (X=C, Si, Ge, Sn) and Blue P; the final MD images for the $SiP_3$ monolayer sheet at 300 and 500 K; Blue P, silicene and $SiP_3$ monolayer optimized structure of possible adsorption sites for single Li, Na and K-ions; $SiP_3$ monolayers and their full alkali-metallized structural parameters; the diffusion barriers of single Li-ion, Na-ion and K-ion on path III, and diffusivity of Li, Na and K-ion along Path I, Path II and Path III on $SiP_3$ monolayer.

■ **AUTHOR INFORMATION**

**Corresponding Authors**

**Changcheng Chen**- *School of Science, Xi'an University of Architecture and Technology, Xi'an, 710055, China; orcid.org/0000-0002-2871-4510*; Email: chenchangcheng@xauat.edu.cn

**Liyuan Wu**- *CAS Key Laboratory for Biomedical Effects of Nanomaterials and Nanosafety, Institute of High Energy Physics, Chinese Academy of Sciences, Beijing 100049, P. R. China;* Email: wuly@bupt.edu.cn

**Pengfei Lu**- *State Key Laboratory of Information Photonics and Optical Communications, Beijing University of Posts and Telecommunications, Beijing, 100876, China;* Email: photon.bupt@gmail.com

**Authors**

**Yue Kuai**- *School of Science, Xi'an University of Architecture and Technology, Xi'an, 710055, China*

**Shuli Gao**- *School of Science, Xi'an University of Architecture and Technology, Xi'an, 710055, China*

**Wen Chen**- *School of Science, Xi'an University of Architecture and Technology, Xi'an, 710055, China*

**Jinbo Hao**- *School of Science, Xi'an University of Architecture and Technology, Xi'an, 710055, China*

**Ge Wu**- *School of Science, Xi'an University of Architecture and Technology, Xi'an, 710055, China*

**Feng Chen-** *School of Science, Xi'an University of Architecture and Technology, Xi'an, 710055, China*

**Shuangna Guo-** *School of Science, Xi'an University of Architecture and Technology, Xi'an, 710055, China*

**Notes**

There are no conflicts to declare.

■ **ACKNOWLEDGMENTS**

This work was supported by the Open-Foundation of Key Laboratory of Laser Device Technology, China North Industries Group Corporation Limited (No. KLLDT202001), the Natural Science Foundation of Shaanxi Province (No. 2021JM-371), and the Fund of State Key Laboratory of IPOC(BUPT) (No. IPOC2019A013).